\documentclass{aa}  

\usepackage{graphicx}
\usepackage{txfonts}
\usepackage{lscape}
\usepackage{tabularx}
\usepackage{adjustbox}
\usepackage{comment}
\usepackage{hyperref}
\begin{document} 
\newcommand{\deufrac}{\smash{$D_{\rm frac}^{\rm N_2H^+}$}}

   \title{Interaction between the Supernova Remnant W44 and the Infrared Dark Cloud G034.77-00.55: shock induced star formation?}


   \author{G. Cosentino\inst{1}\thanks{E-mail:giuliana.cosentino@eso.org},
    I. Jim\'{e}nez-Serra\inst{2},
    A. T. Barnes\inst{1},
    J. C. Tan\inst{3,4}\fnmsep,
    F. Fontani\inst{5},
    P. Caselli\inst{6},
    J. D. Henshaw\inst{7},
    C.-Y. Law\inst{5},\\
    S. Viti\inst{8},
    R. Fedriani\inst{9},
    C.-J. Hsu\inst{3},
    P. Gorai\inst{10,11},
    S. Zeng\inst{12},
    M. De Simone\inst{1,5}}
    \authorrunning{Cosentino et al.}
    \titlerunning{The W44-G034.77 interaction: shock induced star formation?}
    \institute{European Southern Observatory, Karl-Schwarzschild-Strasse 2, D-85748 Garching, Germany
    \and 
    Centro de Astrobiolog\'{i}a (CSIC/INTA), Ctra. de Torrej\'on a Ajalvir km 4, Madrid, Spain
   \and 
   Department of Space, Earth and Environment, Chalmers University of Technology, SE-412 96 Gothenburg, Sweden
    \and 
    Department of Astronomy, University of Virginia, 530 McCormick Road Charlottesville, 22904-4325 USA
    \and
    INAF  Osservatorio Astronomico di Arcetri, Largo E. Fermi 5, 50125 Florence, Italy
    \and
    Max Planck Institute for Extraterrestrial Physics, Giessenbachstrasse 1, 85748 Garching bei M\"{u}nchen, Germany
    \and
    Astrophysics Research Institute, Liverpool John Moores University, 146 Brownlow Hill, Liverpool L3 5RF, UK
    \and
    Leiden Observatory, Leiden University, PO Box 9513, 2300 RA Leiden, The Netherlands
    \and
    Instituto de Astrof\'isica de Andaluc\'ia, CSIC, Glorieta de la Astronom\'ia s/n, E-18008 Granada, Spain
    \and
    Rosseland Centre for Solar Physics, University of Oslo, PO Box 1029 Blindern, 0315 Oslo, Norway
    \and
    Institute of Theoretical Astrophysics, University of Oslo, PO Box 1029 Blindern, 0315 Oslo, Norway
    \and
    Star and Planet Formation Laboratory, Cluster for Pioneering Research, RIKEN, 2-1 Hirosawa, Wako, Saitama, 351-0198, Japan}
   \date{Received ---; accepted ---}

 
  \abstract
   {How Supernova Remnant (SNR) shocks impact nearby molecular clouds is still poorly observationally constrained. It is unclear if SNRs can positively or negatively affect clouds star formation potential.}
   {We have studied the dense gas morphology and kinematics toward the Infrared Dark Cloud (IRDC) G034.77-00.55, shock-interacting with the SNR W44, to identify evidence of early stage star formation induced by the shock.}
   {We have used high-angular resolution N$_2$H$^+$(1-0) images across G034.77-00.55, obtained with the Atacama Large Millimeter/sub-Millimeter Array. N$_2$H$^+$ is a well known tracer of dense and cold material, optimal to identify gas with the highest potential to harbour star formation.}
   {The N$_2$H$^+$ emission is distributed into two elongated structures, one toward the dense ridge at the edge of the source and one toward the inner cloud. Both elongations are spatially associated with well-defined mass-surface density features. The velocities of the gas in the two structures i.e., 38-41 km s$^{-1}$ and 41-43 km s$^{-1}$ are consistent with the lowest velocities of the J- and C-type parts of the SNR-driven shock, respectively. A third velocity component is present at 43-45.5 km s$^{-1}$. The dense gas shows a fragmented morphology with core-like fragments of scales consistent with the Jeans lengths, masses $\sim$1-20 M$_{\odot}$, densities (n(H$_2$)$\geq$10$^5$ cm$^{-3}$) sufficient to host star formation in free-fall time scales (few 10$^4$ yr) and with virial parameters that hint toward possible collapse.}
   {The W44 driven shock may have swept up the encountered material which is now seen as a dense ridge, almost detached from the main cloud, and an elongation within the inner cloud, well constrained in both N$_2$H$^+$ emission and mass surface density. This shock compressed material may have then fragmented into cores that are either in a starless or pre-stellar stage. Additional observations are needed to confirm this scenario and the nature of the cores.}


   \maketitle
    
%
\section{Introduction}
\begin{figure*}[!htpb]
    \centering
    \includegraphics[width=\textwidth,trim= 0cm 0cm 0cm 0cm, clip=True ]{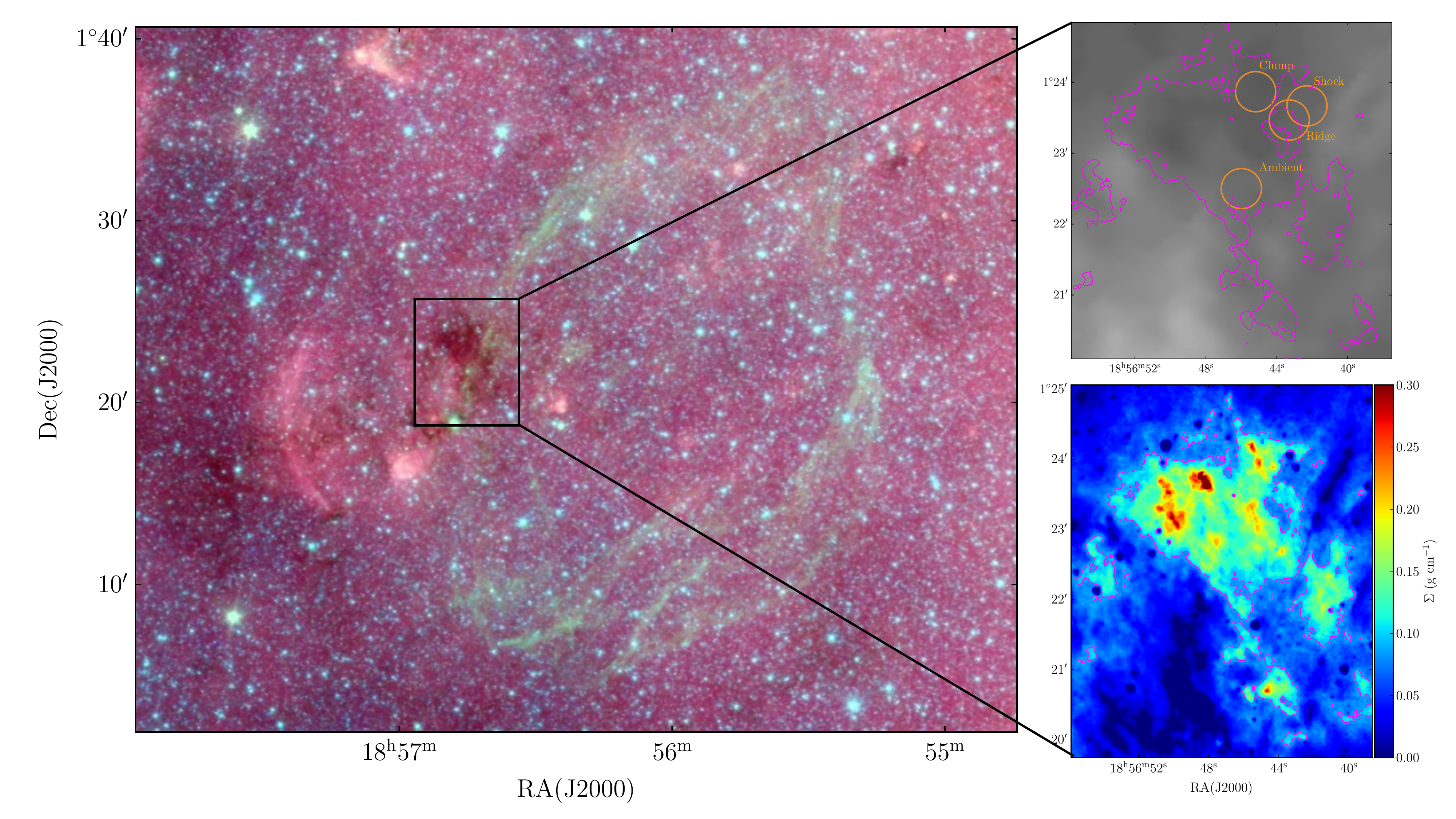}
    \caption{\textit{Left:} Three-color image of G034.77 together with the SNR W44 and the H{\small II} region G034.758-00.681. Red is 8 $\mu$m, green is 5.8 $\mu$m and blue is 4.5 $\mu$m emission from the Spitzer GLIMPSE survey \citep{churchwell2009}. 
    \textit{Right top:} \textit{Hershel} 70 $\mu$m images of G034.77 with overlaid the positions for which the D/H ratios have been obtained in this work (orange circles). The aperture size is 34$^{\prime\prime}$ i.e., the largest aperture in our data. \textit{Right Bottom:} Mass surface density map (colour scale) of G034.77 \citep{kainulainen2013}. In both panels, the $\Sigma$=0.09 g cm$^{-2}$ contour, equivalent to a visual extinction A$_{\rm{V}}$= 20 mag is shown in magenta.}
    \label{fig:Mosaic}
\end{figure*}

It is known that stars are born from the collapse of molecular clouds i.e., the coldest \citep[T$<$25 K;][]{pillai2006,rathborne2006} and densest \citep[n(H$_2$)>10$^{2}$ cm$^{-3}$][]{chevance2023} regions of the Interstellar Medium (ISM). However, the processes that initiate cloud collapse in these objects are still unclear \citep[e.g.,][]{tan2000,tan2014,hernandez2015,peretto2016,RetesRomero2020,morii2021}.
Supernova Explosions (SNes) and Remnants (SNRs) are among the most powerful \citep[energy of detonation $\sim$10$^{50}$-10$^{51}$ ergs;][]{koo2020} events in galaxies and have been proposed to play a pivotal role in setting the low levels of star formation efficiency observed in the local universe \citep[$\epsilon\sim$ 1-2 \%;][]{krumholz2012}. 
\noindent
However, it is still debated whether SNR feedback can efficiently trigger star formation in molecular clouds. For a detailed discussion of the complexity and challenges of this problem, both observationally and theoretically, we refer to \citep{dale15a}. In particular, it is difficult to assess the importance of SNR (and other bubbles) shocks in sweeping up material and hence forming clouds \citep[see e.g. ][]{inutsuka2015} or how pre-existing clouds affect the shock propagation.
Recent observational works seem to indicate that SNR feedback may be able to significantly increase the cloud gas density \cite{inoueFukui2018, cosentino2019,cosentino2022,sano2020,sano2023} 
and the star formation efficiency of galaxies up to 40\% \citep{RicoVillas2020}, but more observational work is needed to inform existing models \citep{ceverino2009,dale2015b}.\\

\noindent
Here, we investigate the dense gas morphology and kinematics toward the Infrared Dark Cloud (IRDC) G034.77-00.55 (thereafter G034.77; Figure~\ref{fig:Mosaic}), known to be shock interacting with the SNR W44 \citep{wootten1978,castelletti2007,cardillo2014}. W44 is a core-collapse SNR of $\sim$20k years and located at a distance of 2.9 kpc \citep{Lee2020}. Today classified as mixed-morphology SNR i.e., with a shell-like morphology in the radio wavelengths and centrally filled in the X-rays \citep{rho1998}, W44 was first identified as a radio source by \cite{Westerhout1958}, then classified as a SNR by \cite{Scheuer1963} and later on observed in X-ray by \cite{Gronenschild1978}. Only more recently \citep{Abdo2010e},  gamma-ray observations with Fermi-LAT were reported. W44 shows a strong elliptical morphology that has long been interpreted as due to the shell expanding into the inhomogeneous interstellar medium (ISM). Indeed, evidence of interaction between the SNR and the surrounding material have been reported at multiple wavelengths \citep[e.g. ][]{Reach2005,Reach2006}. Toward the source North-West, a bright radio emission is associated with atomic line emission characteristic of shock-excited radiative filaments \citep{Giacani1997}. Moreover, toward the same region \cite{Yoshiike2013} used CO emission lines to identify highly-excited molecular gas associated with SNR shell. Toward several regions across the W44, OH maser emission at 1720 MHz has been reported, which is known to indicate sites of ongoing C-type shock interactions \citep{Hoffman2005}. Finally, toward the south-east, W44 is known to be expanding into the dense molecular cloud G034.77. The cloud was first identified by \citep{wootten1978} by CO observations and later on classified as Infrared Dark Cloud by \cite{simon2006} using the Midcourse Space Experiment. The authors also reported the cloud to be located at a distance of 2.9 kpc i.e., consistent with that inferred for W44. Since then, the cloud has been largely studied as a stand-alone source, part of the large sample of \cite{rathborne2006} but also as part of the \cite{butlerTan2009} sample \citep[e.g. ][]{hernandez2015}. The interaction between the cloud and the SNR has also been  investigated. \cite{castelletti2007} reported strong evidence of interaction using VLA radio emission, while \cite{cardillo2014} reported evidence of cosmic ray acceleration. \cite{sashida2013} used HCO$^+$ single-dish maps to infer a shell expansion velocity of $\sim$13 km s$^{-1}$. Thanks to its relative closeness, its relatively mid-age and slow shock, as well as its clear association with a dense cloud, W44 is among the best galactic candidate to study in detail the effect of the interactions between SNRs and molecular clouds. Recently, we have used ALMA images of the J=2-1 Silicon Monoxide (SiO) emission to study the shock interaction between the cloud and the SNR \citep{cosentino2018,cosentino2019}. SiO is a unique shock tracer, because its abundance is usually very low in quiescent regions but that is extremely enhanced in the presence of shocks, where dust sputtering occurs \citep{martinpintado1992,jimenezserra2005}. From the high-angular resolution ALMA images, the SiO emission is organised into two elongated structures caught in the act of decelerating (from $\sim$46 km s$^{-1}$ to $\sim$39 km s$^{-1}$) and plunging into a dense ridge at the edge of G034.77. The plunging shock enhances the ridge gas density by a factor $>$10 i.e., to values n(H$_2$)$>$10$^5$ cm$^{-3}$, compatible to those required to enable star formation \citep{cosentino2019,parmentier2011}. From \cite{cosentino2019}, the SiO kinematic structure is best reproduced by a CJ-type time-dependent magneto-hydro-dynamic shock propagating into the cloud at a velocity of $\sim$ 20 km s$^{-1}$, consistent with that measured by \cite{sashida2013} using HCO$^+$(1-0) observations.\\ 
More recently, we used IRAM-30m single pointing observations of N$_2$H$^+$(1-0) and N$_2$D$^+$(1-0) to investigate the Deuterium fractionation across the shock front \citep{cosentino2023}. D/H is a well-studied tracer of cores on the verge of gravitational collapse, when is enhanced by several orders of magnitude \cite{crapsi2005,caselli2008,emprechtinger2009} with respect to the cosmic D/H abundance \citep[$\sim$10$^{-5}$;][]{oliveira2003}. Indeed, in these regions, the molecular material is very dense (n(H$_2$)$>$10$^4$ cm$^{-3}$) and cold \cite[T$<$20 K][]{awad2014,fontani2015} and CO is highly depleted from the gas phase. All this boosts the root reaction of many deuterated molecules \citep{gerlich2002}. Toward the dense ridge, the reported D/H ratios lie in the range $\sim$0.05-0.1, consistent with what is typically measured in starless and pre-stellar cores \citep{fontani2011,emprechtinger2009,kong2016,cheng2021}. These values are also a factor $>$2-3 larger than the D/H ratio measured toward an inner region of the cloud and named as Ambient \citep{cosentino2023}. Finally, the deuterated gas velocity is consistent with that of the post-shocked gas probed by C$^{18}$O \citep{cosentino2019}, suggesting that the starless and pre-stellar cores may have formed by the interaction of the shock.\\

\noindent
In this work, we complement the IRAM-30m spectra presented in \cite{cosentino2023}, with APEX observations of the J=3$\rightarrow$2 N$_2$H$^+$ and N$_2$D$^+$ transitions and use a multi-line approach to estimate the D/H ratio across the shock front. Hence, we use the N$_2$H$^+$(1-0) ALMA images from \cite{barnes2021} to investigate the morphology and kinematics of the dense gas across and near the shock. N$_2$H$^+$ is a well known tracer of dense and cold material, where  its destroyer CO depletes much faster than its precursor N$_2$ \citep{caselli2002}. Therefore, N$_2$H$^+$ has also been widely used to identify cold cores in molecular clouds \citep[e.g. ][]{pirogov2003,tatematsu2004,fontani2006,liu2019}
The paper is organised as follows. In Sect. ~\ref{ObsAndData}, we describe the new observations and archival data used in this work. In Sect. ~\ref{DoverH}, we obtain new, more accurate, estimates of  the D/H ratios across the shock front. In Sect. ~\ref{N2Hpanalysis} and ~\ref{CoresAnalysis}, we investigate the N$_2$H$^+$(1-0) morphology and kinematics, using the archival ALMA images. Finally, in Sect. ~\ref{DiscAndConcs}, we discuss our results and present our conclusions. 

\section{Observations and data}\label{ObsAndData}
\subsection{APEX and IRAM-30m observations}
In March 2020, we obtained high-sensitivity single pointing observations of the N$_2$H$^+$ and N$_2$D$^+$ J=1-0 transitions toward five positions across the IRDC G034.77 \citep{cosentino2023}. For a complete description of the IRAM-30m observations we refer to \cite{cosentino2023}. The five positions were chosen to be representative of different physical conditions across the cloud. In particular, the positions labelled as 'Clump', 'Ridge' and 'Shock' are located toward the shock front probed by the SiO emission \citep{cosentino2019} and are representative of a dense clump adjacent to the shock front, the dense ridge and the SiO emission peak \cite[see Figure 1 in ][]{cosentino2018}. Toward these three positions, in July 2023 we used Atacama Path-Finder Experiment (APEX) to obtained high-sensitivity single pointing observations of the N$_2$H$^+$ and N$_2$D$^+$ J=3-2 transitions. In Figure~\ref{fig:Mosaic} (top left panel), the three positions of interest are shown (orange circle), together with the location of the Ambient region. In \cite{cosentino2023}, the Ambient position was selected for being representative of the bulk of the cloud i.e., the inner gas within G034.77 with velocity $\sim$43 km s$^{-1}$. For the three positions 'Clump', 'Ridge' and 'Shock', observations were performed in position switching mode (off-position RA(J2000)=18$^h$57$^m$01$^s$, Dec(J2000)=1$^d$22$^m$25$^s$) and dual polarisation mode. The nFLASH240 receiver was used with tuning frequencies 231.320 and 279.512 GHz and frequency resolution 488 kHz. At the selected tuning frequencies, the velocity resolutions and beam sizes are 0.7 km s$^{-1}$ and 27$^{\prime\prime}$ for N$_2$D$^+$(3-2), 0.5 km s$^{-1}$ and 23$^{\prime\prime}$ for N$_2$H$^+$(3-2). For comparison, the IRAM-30m spectra presented in \cite{cosentino2023} have spectral resolution $\sim$0.7 km s$^{-1}$ and beam sizes 34$^{\prime\prime}$ and 27$^{\prime\prime}$ for N$_2$D$^+$(1-0) and N$_2$H$^+$(1-0), respectively. Observations were obtained in unit of antenna temperature, T$^*_A$, and converted to main beam temperature T$_{\rm{mb}}$ using beam and forward efficiency of 0.77 and 0.95, respectively. The achieved rms are 3 mK for N$_2$D$^+$(3-2) and 7 mK for N$_2$H$^+$(3-2), while \cite{cosentino2023} report rms of 5 mK for N$_2$D$^+$(1-0) and 7 mK for N$_2$H$^+$(1-0). In Table~\ref{tab:tabAPEX}, the coordinates of the positions of interest are reported. 

\begin{table}[!htpb]
    \caption{Equatorial Coordinates of the positions analysed in this work.}
        \centering

    \begin{tabular}{llllllllllllllllll}
    \hline
     \hline
         Position &RA(J2000) & Dec(J2000)\\
         & (hh:mm:ss) & (dd:mm:ss) \\
         \hline
        Shock   &18:56:42.3 &01:23:40\\
        Ridge   &18:56:43.4 &01:23:28\\
        Clump   &18:56:45.2 &01:23:52\\
    \hline
    \hline
    \end{tabular}
    \label{tab:tabAPEX}
\end{table}

\subsection{ALMA data}
The N$_2$H$^+$(1-0) images here analysed were first presented by \cite{barnes2021} and obtained as part of a high-resolution Atacama Large Millimeter/sub-Millimeter Array (ALMA) survey (project codes 2017.1.00687.S and 2018.1.00850.S, PI: A.T.Barnes) to investigate star formation in the sample of 10 IRDCs first presented by \cite{butlerTan2012}. As reported by the authors, observations were performed using the ALMA Band 3 receiver (frequency $\sim$93 GHz) with velocity resolution 0.1 km s$^{-1}$ and channel width 0.05 km s$^{-1}$. For the observations, the 12m array in configuration C43-1 was used together with the 7m ACA array and the Total Power single-dish. Images from the three configurations were combined using the task {\sc feather} in {\sc casa}. From the final combined data-cube, we extracted spectra toward the regions labelled as Clump, Ridge and Shock, using a circular aperture comparable with the IRAM-30m beam size at the N$_2$H$^+$(1-0) frequency ($\sim$30$^{\prime\prime}$). We thus compared these with the N$_2$H$^+$(1-0) spectra obtained with the IRAM-30m. The three set of spectra are comparable within the rms, indicating that all flux is recovered in the ALMA images within the uncertainty. For a more detailed description of the ALMA set up and reduction process we refer to \cite{barnes2021}. The final images are the same ones used by the authors and have synthesised beam size of 3.05$^{\prime\prime}\times$3.5$^{\prime\prime}$ p.a=-74$^{\circ}$ and rms 0.2 K. In order to increase the signal-to-noise ratio of the map, we have smoothed the cube to have velocity resolution 0.2 km s$^{-1}$.

\subsection{Mass surface density map}
In this work, we also use the mass surface density, $\Sigma$, map obtained toward G034.77-00.55 by \cite{kainulainen2013}. Building on the MIREX method presented by \cite{butlerTan2012}, the authors combined Mid- and Near-Infrared Spitzer images at multiple wavelengths to estimate the mass surface density across the 10 IRDCs sample from \cite{butlerTan2012}. For a detailed description of the MIREX method we refer to \cite{butlerTan2012,kainulainen2013}. The mass surface density map has angular resolution of 2$^{\prime\prime}$, dictated by the resolution of the Spitzer-IRAC maps and a uncertainty of $\sim$30\%. The MIREX method is unable to estimate mass surface density toward sources that are bright at Mid-IR wavelengths, hence the pixels toward which the method fails appear as blank in the map. The $\Sigma$ map is shown in Figure~\ref{fig:Mosaic} (right panel), with superimposed the SiO emission contours obtained in \cite{cosentino2019} and indicated the Ridge, Shock and Clump positions. 

\begin{figure*}[!htpb]
    \centering
    \includegraphics[width=0.95\textwidth, trim=2.5cm 2cm 3cm 3cm, clip=True]{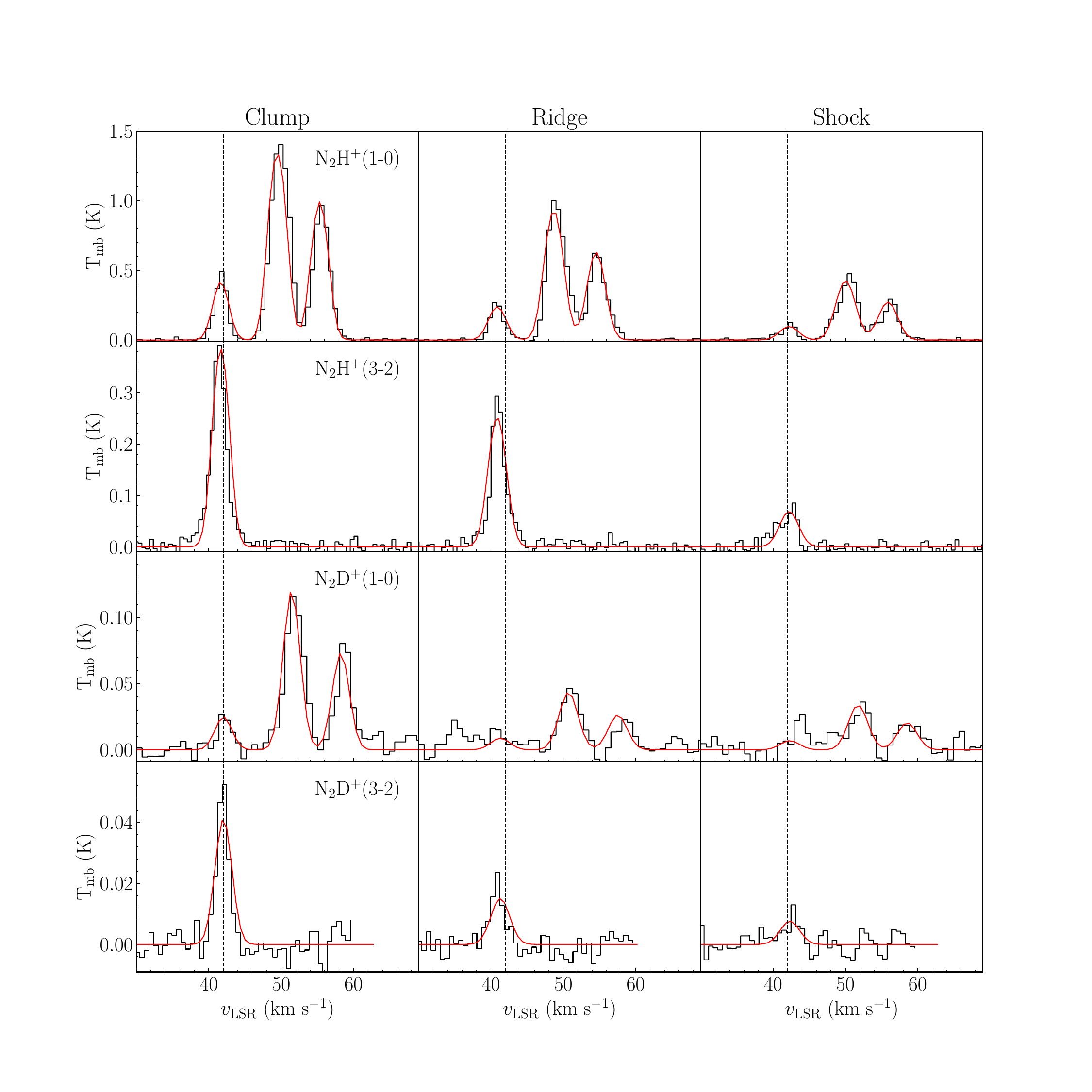}
    \caption{N$_2$H$^+$ (first and third row) and N$_2$D$^+$ (second and fourth row) spectra of the J=1-0 and J=3-2 transitions obtained toward the Clump, Ridge and Shock. The IRDC G34.77 central velocity is indicated as vertical dotted line in all panels. The red curves show the best LTE  models obtained by {\sc madcuba}.}
    \label{fig:spectra}
\end{figure*}

\section{D/H at the W44-G034.77 shock interface: a multi-line approach}\label{DoverH}
In \cite{cosentino2023}, we used the software {\sc madcuba}\footnote{{\sc madcuba} is a software developed in the Madrid Center of Astrobiology (INTA-CSIC). https://cab.inta-csic.es/madcuba/} \citep{martin2019} to obtain Local-Thermodynamic-Equilibrium (LTE) models of the N$_2$H$^+$(1-0) and N$_2$D$^+$(1-0) hyper-fine structures observed with the IRAM-30m. 
In {\sc madcuba}, we used the spectral line identification and local thermodynamic equilibrium modelling (SLIM) tool identify the molecular transitions. Hence, we used the built-in AUTOFIT function of SLIM to obtain the LTE models that best reproduce all transitions. Indeed, AUTOFIT produces the synthetic spectrum that best matches the data, assuming as input parameters the total molecular column density (N$_{\rm{tot}}$), the radial systemic velocity of the source ($v$), the line full width at half maximum (FWHM), the excitation temperature (T$_{\rm{ex}}$), and the angular size of the emission ($\theta_S$). AUTOFIT assumes that $v$, FWHM, $\theta_S$, and T$_{\rm{ex}}$ are the same for all transitions that are fitted simultaneously. We set the angular size of the emission, $\theta_S$, to reflect the assumption that the emission fills the beam i.e., filling factor of 1. As it will be shown, this assumption holds for N$_2$H$^+$(1-0) toward the Clump and Ridge but may not be the case for the Shock position. However, with the current data in hand we cannot assess this for the additional transitions/species. We thus maintain the assumption of emission filling the beam for all transitions.\\ In \cite{cosentino2023}, we assumed T$_{\rm{ex}}$ and used {\sc madcuba} to obtain the N$_2$H$^+$ and N$_2$D$^+$ column densities, N(N$_2$H$^+$) and N(N$_2$D$^+$), toward the different regions. Hence, we estimated the deuterium fractionation as:

\begin{equation}
    D_{\rm frac}^{\rm N_2H^+} = \frac{N({\rm N_2D^+})}{N({\rm N_2H^+})},
    \label{eq1}
\end{equation}

\noindent
Toward the three positions, we found $D_{\rm frac}^{\rm N_2H^+}\sim$0.05-0.1 i.e., enhanced by a factor $\sim$2-3 with respect to the $D_{\rm frac}^{\rm N_2H^+}$ 3 $\sigma$ upper limit ($<$0.03) estimated toward the Ambient region i.e., the unperturbed gas where neither shock or star formation activity are present.\\ Since a single rotational line was available, we assumed excitation temperature T$_{\rm{ex}}$=9 K for both species and treated this as a fixed parameter in {\sc madcuba}. This value was obtained by \cite{cosentino2018} from the LTE analysis of multiple CH$_3$OH transitions at 3 mm toward G034.77. We now complement the IRAM-30m data with APEX single pointing observations of the N$_2$H$^+$ and N$_2$D$^+$ J=3$\rightarrow$2 transitions, toward Clump, Ridge and Shock.\\ 
From the advantage of a multi-transitions approach, we consider the species excitation temperatures as free parameters and use {\sc madcuba} to obtain the best LTE models that simultaneously reproduce the J=1-0 and J=3-2 transitions. The J=1-0 spectra from \cite{cosentino2023}, the new J=3-2 spectra and the best {\sc madcuba} LTE models are shown in Figure~\ref{fig:spectra}. The best model parameters are also reported in Table~\ref{tab:tab1}, together with the values already obtained in \cite{cosentino2023}. The uncertainties given by {\sc madcuba} are $<$10\% for T$_{\rm{ex}}$ and $\sim$2\% for $v_{\rm{0}}$ and $\Delta v$. The final uncertainty on the last two quantities is obtained by summing in quadrature the {\sc madcuba} error with the spectra channel width (0.6-0.8 km s$^{-1}$ or $\sim$10\%). Finally, the uncertainty on \deufrac is obtained by adding in quadrature the uncertainties on $N$(N$_2$H$^+$) and $N$(N$_2$D$^+$).\\

\begin{table*}[!htpb]
    \caption{N$_2$H$^+$ and N$_2$D$^+$ best LTE model parameters as obtained by {\sc madcuba} (odd lines) toward the three positions and corresponding \deufrac. }
    \centering
    \begin{tabular}{ccccc|cccc|cc}
    \hline
     \hline
& \multicolumn{4}{c}{N$_2$H$^+$} & \multicolumn{4}{c}{N$_2$D$^+$} & \\
& $N_{tot}$  & T$_{\rm{ex}}$ & $v_{\rm{0}}$ & $\Delta v$ & $N_{tot}$ & T$_{\rm{ex}}$ & $v_{\rm{0}}$ & $\Delta v$ &\deufrac & Reference\\
&(10$^{13}$ cm$^{-2}$) & (K) & (km s$^{-1}$) & (km s$^{-1}$)& (10$^{12}$ cm$^{-2}$) & (K) & (km s$^{-1}$) & (km s$^{-1}$) & \\
      \hline
Shock & 0.50$\pm$0.04 & 4.4 & 42.2 &3.0 & 0.5$\pm$0.1 & 3.9 & 42.3 &3.1 & 0.080$\pm$0.006 & This work\\
       & 0.36$\pm$0.09 & 9 &42.2 &3.2 & 0.3$\pm$0.1 & 9 &  42.2 & 3.2 & 0.07$\pm$0.03 & Cosentino+23\\
      Ridge & 1.10$\pm$0.05 & 4.9 & 40.9 &2.8 & 0.6$\pm$0.07 & 4.4 & 41.3 &3.1 &0.055$\pm$0.007 & This work\\
       &0.8$\pm$0.2 & 9 & 40.9 & 3.1 & 0.4$\pm$0.1 &9 & 41.4 & 3.1 & 0.05$\pm$0.03 & Cosentino+23\\
      Clump & 1.70$\pm$0.08  & 4.8 & 41.7 &2.5 & 1.4$\pm$0.08 & 4.3 & 42.0 &2.7 &0.10$\pm$0.02 & This work\\
       & 1.1$\pm$0.1 & 9 & 41.7 & 2.8 & 1.2$\pm$0.1 &9 & 42.1 & 3.0 & 0.10$\pm$0.02 & Cosentino+23\\
    \hline
    \hline
    \end{tabular}
    \tablefoot{For each position, we also report the corresponding values already presented in \cite{cosentino2023} (Cosentino+23; even lines).}
    \label{tab:tab1}
\end{table*}

\noindent
From Table~\ref{tab:tab1}, the T$_{\rm{ex}}$ here obtained are a factor of 2 lower that that assumed in \cite{cosentino2023} but consistent between the two species. As a consequence, the column densities and therefore the \deufrac here obtained are in agreement with those inferred in \cite{cosentino2023}. The 3$\sigma$ D/H upper limit previously estimated for the Ambient region does not change significantly when T$_{\rm{ex}}$=4.5 K is assumed. For all species and toward all positions the optical depth, $\tau$ is $<<1$. The multi-transition approach here adopted confirms the results reported in \cite{cosentino2023} and its implications. In particular, the \deufrac values across the shock front between W44 and G034.77 are consistent with those typically measured in dense and cold material with high-potential to yield star formation, e.g. both low- and high-mass pre-stellar and starless cores \citep{emprechtinger2009,fontani2011,friesen2013,kong2016,punanova2016}. 

\section{N$_2$H$^+$ kinematics and morphology: fragmented vs continuous material.}\label{N2Hpanalysis}
\begin{figure}
    \centering
    \includegraphics[width=0.5\textwidth,trim = 1cm 2cm 0cm 4cm, clip=True]{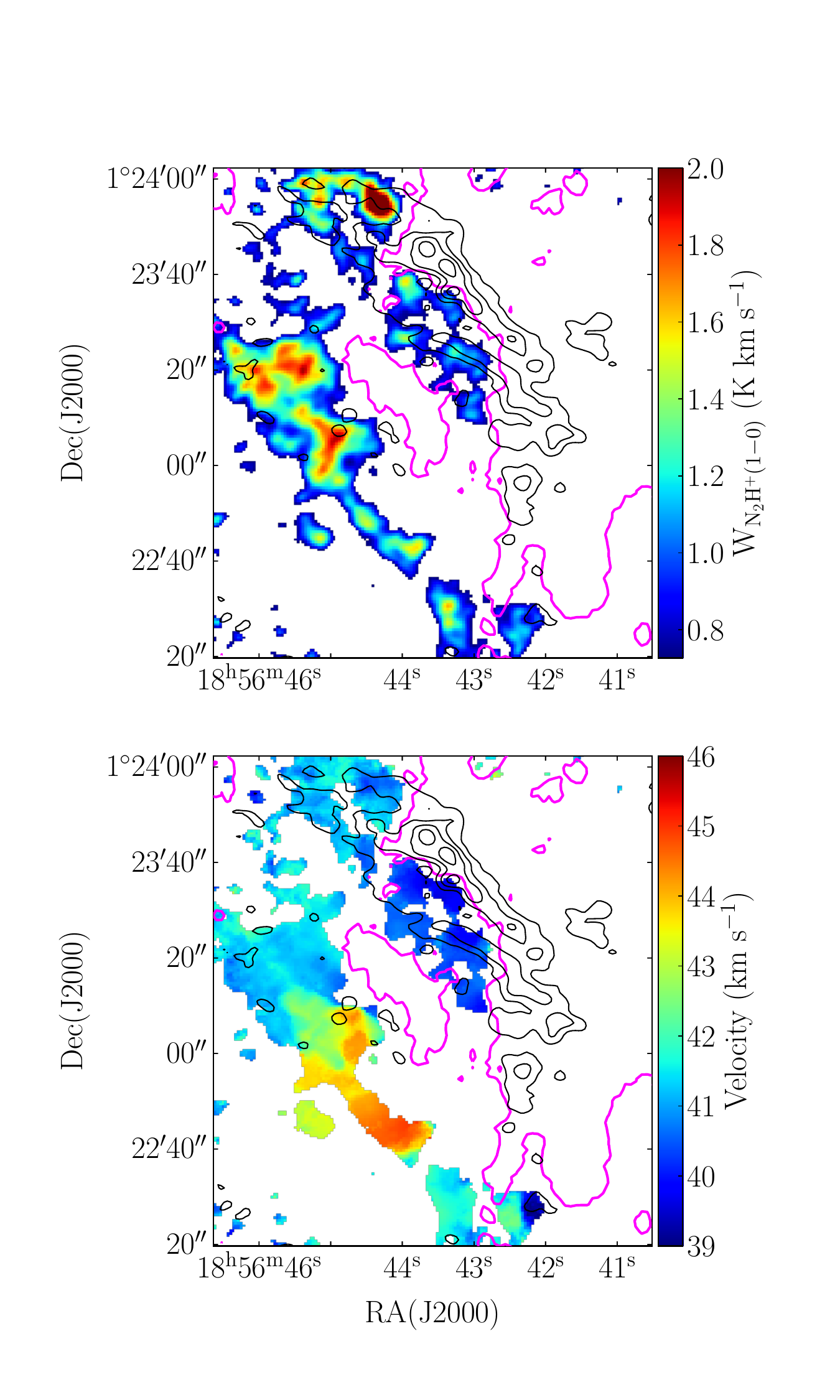}
    \caption{Integrated intensity map (top panel) and velocity field map (bottom panel) of the N$_2$H$^+$(1-0) emission toward the shock front in G034.77. In both panels, pixels below 3$\times$A$_{\rm{rms}}$ have been masked (Arms= 0.23 K km s$^{-1}$). The $\Sigma_{H_2}$=0.09 g cm$^{-2}$ (visual extinction 20 mag) magenta contour highlights the shape of the cloud \citep{kainulainen2013}. The SiO(2-1) emission contours obtained by \cite{cosentino2019} is shown as black contours from 3$\sigma$ ($\sigma$=0.016 Jy beam$^{-1}$ km s$^{-1}$) by steps of 3$\sigma$. }
    \label{fig:fig2}
\end{figure}

We now use the high-angular resolution ALMA images of the N$_2$H$^+$(1-0) emission toward G034.77 \citep{barnes2021,fontani2021} to investigate the morphology and kinematics of the dense gas in the proximity of the shock front. The N$_2$H$^+$ integrated intensity map (39-45.5 km s$^{-1}$; top panel) is shown in Figure~\ref{fig:fig2}, together with the velocity field (moment 1) map (bottom panel). In both panels, the SiO emission contours are overlaid (black contours) and the pixels below 3$\times$A$_{rms}$\footnote{A$_{\rm{rms}}$=rms$\times dv\times\sqrt{\rm{N_{channels}}}$, where rms is the spectra noise, $dv$ is the spectral resolution in km s$^{-1}$ and N$_{\rm{channels}}$ is the number of channels that have been integrated} have been masked. The $\Sigma_{H_2}$=0.09 g cm$^{-2}$ (visual extinction 20 mag) \citep{kainulainen2013} contour (magenta) highlights the shape of the cloud. We note that all the integrated intensity maps presented in this work are obtained considering only the emission from the isolated component of the N$_2$H$^+$ hyper-fine structure, which is optically thin. 

\noindent
As shown in Figure~\ref{fig:fig2}, the dense gas is spatially distributed across the ridge and further in, within the cloud, but it is not widespread everywhere toward the source. The emission is indeed mostly distributed along two parallel elongations that notably are also parallel to the direction of the shock front probed by the SiO emission. Both elongations show multiple emission peaks, the brightest of which is located in correspondence of the region named as the Clump in \cite{cosentino2023}. From the velocity field map (right panel), the dense gas emission across the ridge appears as a velocity-coherent component at $\sim$39-41 km s$^{-1}$. On the other hand, the elongation within the cloud appears as the superposition along the line of sight of at least two velocity-coherent components i.e., a central, more compact structure with velocity $\sim$43-45 km s$^{-1}$ and a more widespread, elongated emission at $\sim$41-42 km s$^{-1}$.\\

\begin{figure}
    \centering
    \includegraphics[width=0.5\textwidth]{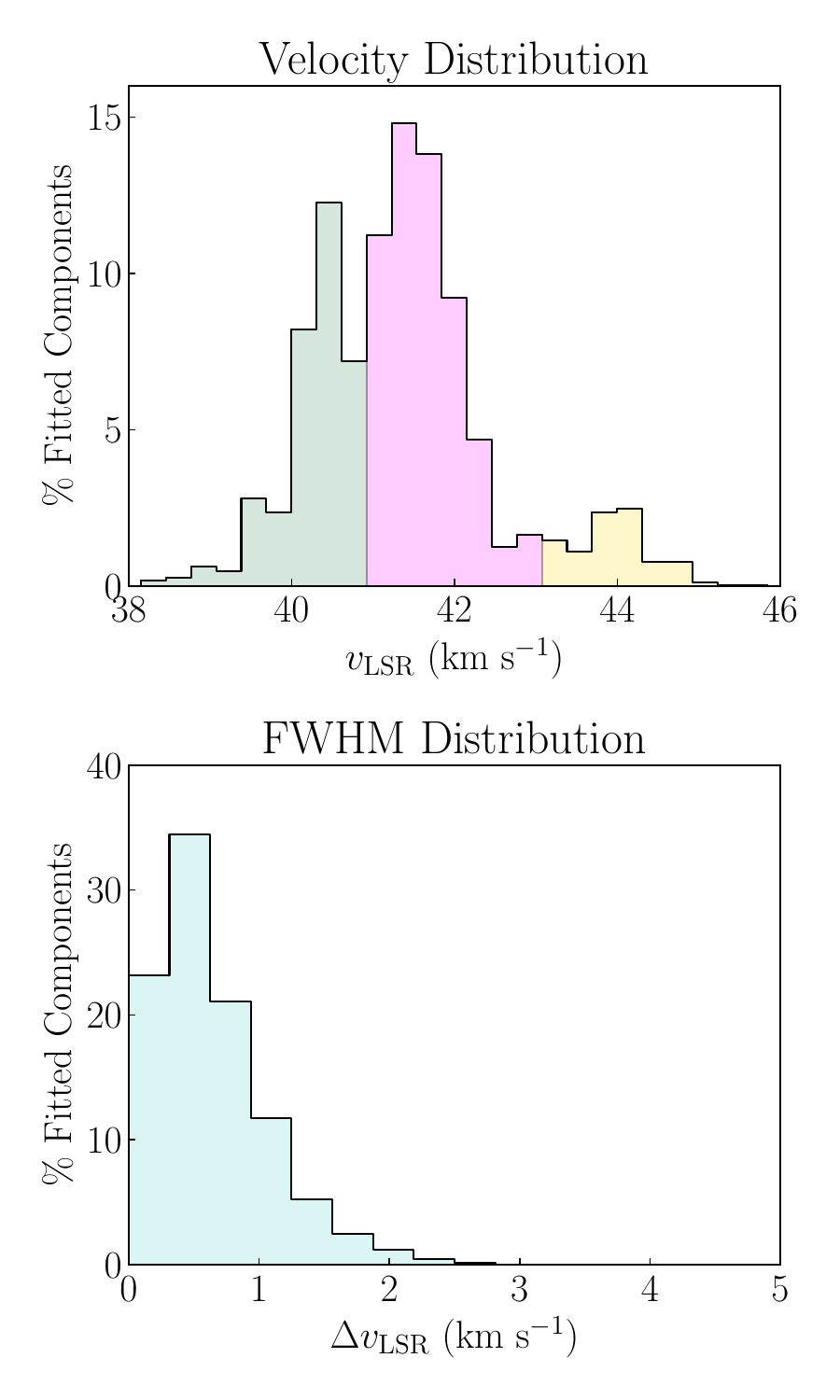}
    \caption{Velocity (top panel) and FWHM (bottom panel) distributions of the N$_2$H$^+$(1-0) emission toward G034.77. Bin sizes are 0.3 km s$^{-1}$ which corresponds to 1/3 of mean intensity-weighted linewidth.}
    \label{fig:figSCOUSE}
\end{figure}

\noindent
In order to more accurately constrain these components, we have used the semi-automated {\sc python} tool {\sc scousepy} \cite{henshaw2016} to perform multi-Gaussian fitting of all the spectra in the data cube. The {\sc scousepy} output provides information on the velocity, intensity and FWHM of the multiple velocity components toward each positional element in a data cube. We have obtained this information for the N$_2$H$^+$ emission mapped with ALMA ($\sim$30000 spectra fitted) and used it to build velocity (top panel) and FWHM (bottom) distributions for the dense gas emission in G034.77. For a detailed description of {\sc scousepy}, we report to \cite{henshaw2016,henshaw2019}. For the {\sc scousepy} analysis of the ALMA N$_2$H$^+$ emission, we have re-binned the data to have velocity channel 0.2 km s$^{-1}$ and have considered only the isolated component of the N$_2$H$^+$(1-0) hyper-fine structure. The velocity and FWHM distributions built from the {\sc scousepy} output are shown in Figure~\ref{fig:figSCOUSE}. From the FWHM distribution, the dense gas emission is on overall relatively narrow (mean intensity weighted FWHM = 0.9 km s$^{-1}$), with line widths narrower than 2 km s$^{-1}$. As already hinted by the velocity field map, the velocity distribution in Figure~\ref{fig:figSCOUSE} indicates the presence of multiple velocity component in the dense gas: a component spatially associated with the dense ridge, with velocities between $\sim$38-41 km s$^{-1}$, ii) a component located within the inner cloud, with velocities in the range $\sim$41-43 km s$^{-1}$ and iii) a third, higher-velocity component ($\sim$43-45 km s$^{-1}$) also located within the cloud but showing a more compact morphology.\\

\begin{figure}
    \centering
    \includegraphics[width=0.47\textwidth,trim = 0cm 3.5cm 0cm 5cm, clip=True]{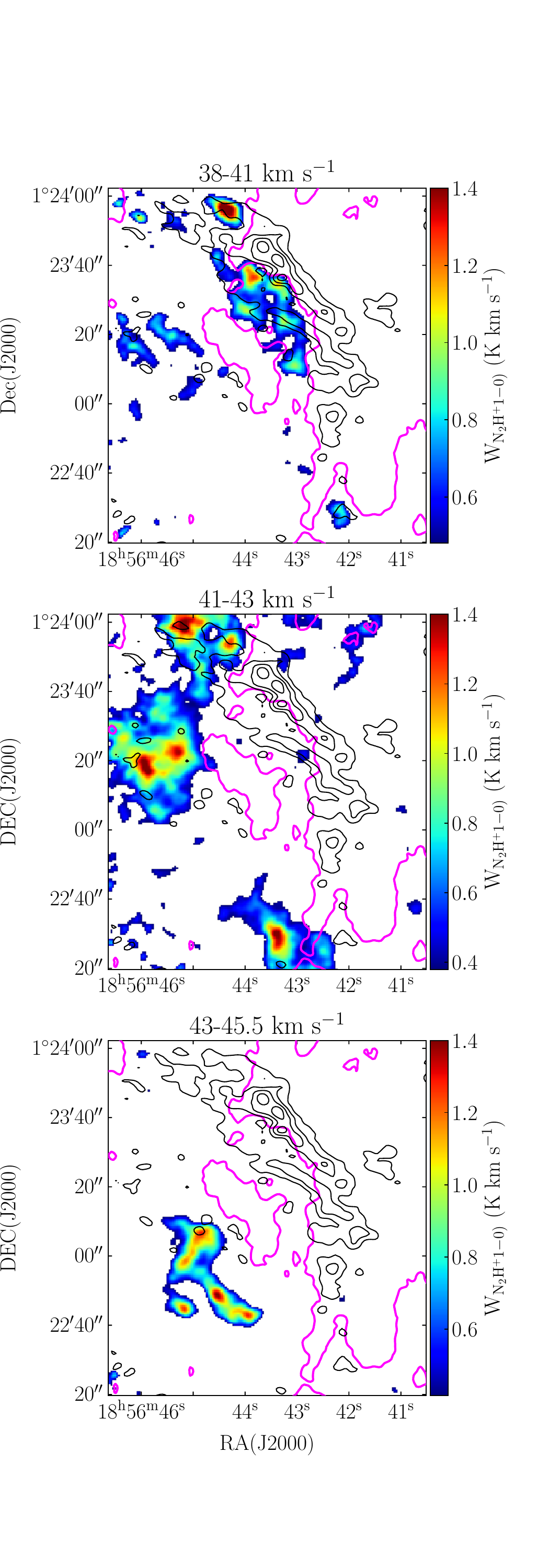}
    \caption{Integrated intensity maps of the N$_2$H$^+$(1-0) emission for the three identified velocity components i.e., 38-41 km s$^{-1}$ (top panel; A$_{\mathrm{rms}}$=0.16 K), 41-43 km s$^{-1}$ (middle panel; A$_{\mathrm{rms}}$=0.13 K) and 43-45.5 km s$^{-1}$ (bottom panel; A$_{\mathrm{rms}}$=0.14 K). In all panels, pixels below 3$\times$A$_{\rm{rms}}$ have been masked. The $\Sigma_{H_2}$=0.09 g cm$^{-2}$ (visual extinction 20 mag) magenta contour highlights the shape of the cloud \citep{kainulainen2013}. The SiO(2-1) emission \citep{cosentino2019} is shown as black contours from 3$\sigma$ ($\sigma$=0.016 Jy beam$^{-1}$ km s$^{-1}$) by steps of 3$\sigma$.}
    \label{fig:figFils}
\end{figure}

In order to better identify the spatial distribution of each velocity component, we have obtained integrated intensity maps toward the three velocity ranges mentioned above, as shown in Figure~\ref{fig:figFils}. 

\noindent
From Figure~\ref{fig:figFils}, the three components are very well spatially separated. The lower velocity component (38-41 km s$^{-1}$) is mostly located toward the dense ridge and well follows its shape, while the other two components are located toward the inner cloud. A partial overlap of the 38-41 km s$^{-1}$ and 41-43 km s$^{-1}$ components is found toward the north and east of the cloud (with east being the direction of increasing RA(J2000)), where the region labelled as Clump is located. Finally, the spatial distribution of the velocity component at 43-45.5 km s$^{-1}$ appears as complementary to the 41-43 km s$^{-1}$ component and it is relatively compact and spatially corresponding to a bright feature (i.e., density enhancement) in the $\Sigma$ map. From Figure~\ref{fig:figFils}, all three components show several peaks in the N$_2$H$^+$ emission. However, while the emission peaks at 41-45.5 km s$^{-1}$ all lie on top of a more extended and fainter emission, the emission peaks at velocities 38-41 km s$^{-1}$ appear as little fragments or core-like structures almost detached from each other. We now investigate the nature of these fragments, and infer their physical and dynamical properties.

\section{Shock induced star formation in G034.77}\label{CoresAnalysis}
\subsection{Core identification.}
\begin{figure}
    \centering
    \includegraphics[width=0.5\textwidth,trim=0cm 0cm 0cm 0cm, clip=True]{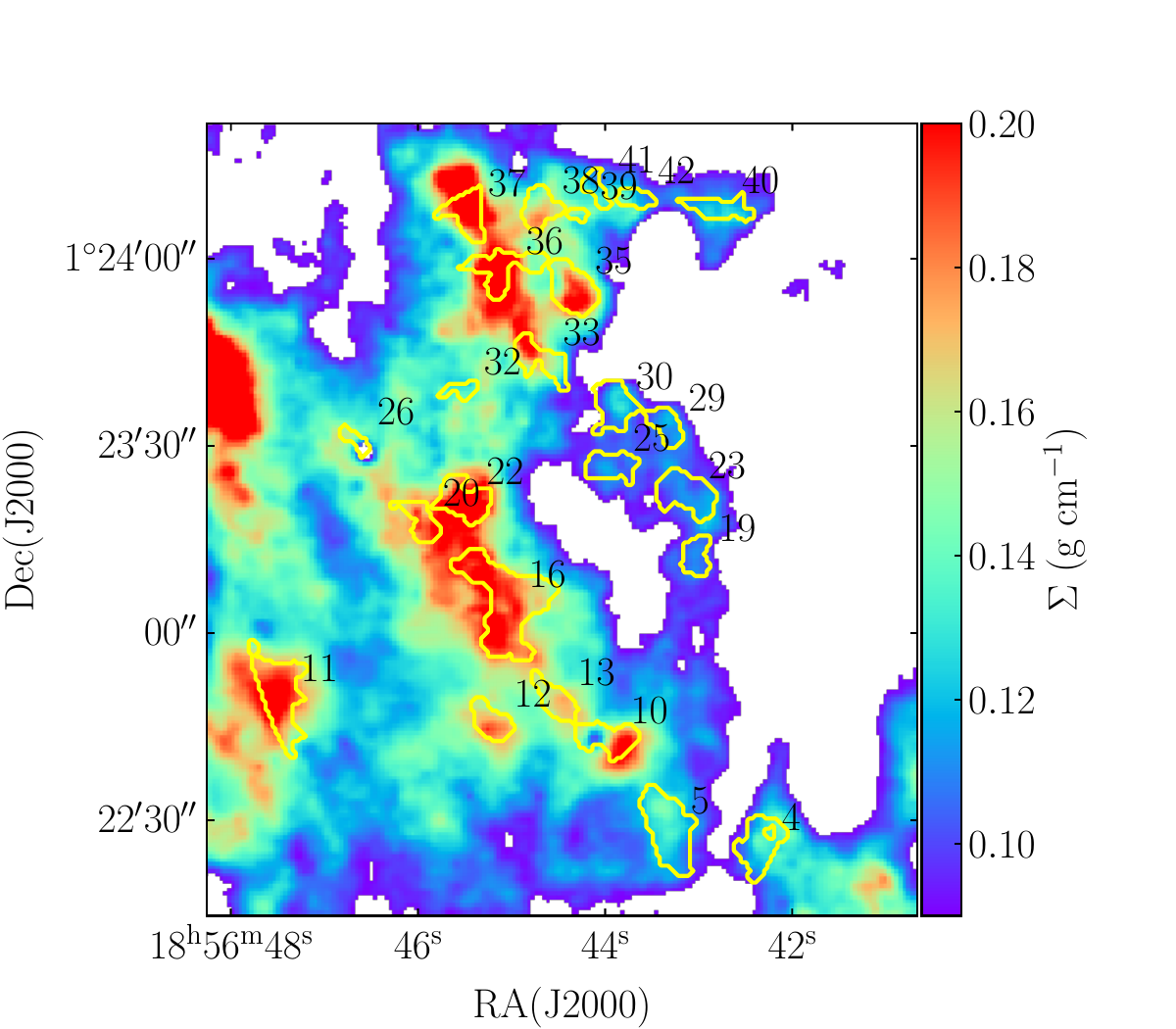}
    \caption{Position of the cores (yellow) identified by the dendogram analysis and overlaid mass surface density map from \cite{kainulainen2013} (color scale). Pixels below a threshold of $\sigma$=0.09 g cm$^{-2}$ have been masked.}
    \label{fig:FigDendo}
\end{figure}

In order to systematically constrain the position and extent of the fragment-like structures, we now perform a dendogram analysis \citep{rosolowsky2008} on the full (38-45.5 km s$^{-1}$) N$_2$H$^+$ integrated intensity map. Dendograms provide a robust and consistent method for the identification of hierarchical structures within maps and cubes. The smallest of these structures are called leaves and have been widely considered good identifier of core-like objects. We build dendogram using the {\sc python} package {\sc astrodendro}\footnote{\url{https://dendrograms.readthedocs.io/en/stable/}}, which requires the user to define a set of parameters. These are the {\sc min$\_$value} (the minimum intensity to be considered by the algorithm), the {\sc min$\_$delta} (the minimum intensity difference between isocontours) and {\sc min$\_$pix} (the minimum number of pixels within a structure). Following the analysis of \cite{barnes2021} and the method adopted in several previous works \citep{cheng2018,liu2018,oneill2021}, we set {\sc min$\_$value}=3$\sigma$ = 0.70 K km s$^{-1}$, {\sc min$\_$delta}= 1$\sigma$ = 0.23 K km s$^{-1}$ and {\sc min$\_$pix}= 0.5 beam = 18 pixels and perform the analysis on the non primary beam corrected image. However, since most of the region of interest is located toward the centre of the map, the analysis performed on the primary beam corrected image is not significantly different. From the dendogram analysis on the N$_2$H$^+$ integrated intensity map we identify 41 leaves. From this list, we exclude all those leaves for which the extracted spectra show no significant emission i.e., they are artefacts due to noise effects and are mostly located at the edge of the map. The final number of significant leaves is 26 and we will refer to these as cores. In Table~\ref{tab:TabCores}, we report the physical and dynamical properties obtained for the cores here identified. In Figure~\ref{fig:FigDendo}, we show the 26 cores as overlaid on the mass surface density ($\Sigma$). 

\noindent
From Figure~\ref{fig:FigDendo}, several cores are identified throughout the source, and especially within the ridge. The mass surface density peak located in correspondence of the region named Clump is also clearly seen together with two additional cores nearby. The majority of the identified cores are spatially associated with well defined peaks in the mass surface density map (Figure~\ref{fig:FigDendo}). This is particularly clear toward the ridge, where the emission is not saturated due to the bulk of the $\Sigma$ emission from the main cloud. From the $\Sigma$ map, we estimate the core mass surface densities as the average $\Sigma$ value within the corresponding leaf and covert this to mass by multiplying for the leaf area. We obtain masses in the range 1-20 M$_{\odot}$. In particular, the cores in correspondence of the Clump region has a mass of $\sim$10 M$_{\odot}$. Since the MIREX method used to obtain the $\Sigma$ map are known to slightly underestimate the mass surface density in bright continuum regions \citep{kainulainen2013}, the obtained masses are likely lower limits. Assuming spherical geometry for the cores, we estimate their volume density as:

\begin{equation}
    n_{\rm{H_2}} = \frac{M}{\frac{4}{3} \pi r^3 \mu_{\rm H_2} m_{\rm{H}}}
\end{equation}

\noindent
where $\mu_{\rm H_2}$=2.8 is the mean molecular weight per hydrogen molecule and m$_{\rm{H}}$ is the mass of the hydrogen atom. $r$ is the effective radius of the core. We find volume densities in the range (4-20)$\times$10$^5$ cm$^{-3}$. The corresponding free-fall times are in the range (2-5)$\times$10$^4$ yr and have been estimated as follows:

\begin{equation}
    t_{\rm{ff}} = \Bigg(\frac{\pi^2 r^3}{8 G M}\Bigg)^{0.5}
\end{equation}

\noindent
where G is the gravitational constant.\\

\subsection{Core stability against thermal, turbulent and magnetic support.}
In order to investigate whether cores are stable against thermal support, we estimate their Jeans masses and lengths as follows:

\begin{equation}
    M_{\rm{J}} = \frac{\pi^{5/2} c_{\rm{s}}^{3} } {6 G^{3/2} \rho^{1/2}}
\end{equation}

\begin{equation}
    \lambda_{\rm{J}} = c_{\rm{s}} \Bigg( \frac{\pi}{G \rho}\Bigg)^{1/2}
\end{equation}

\noindent
where $\rho$ is the core volume mass density and c$_{\rm{s}}$ is the sound speed, defined as:

\begin{equation}
    c_{\rm{s}} = \sqrt{\frac{k_{\rm{b}} T_{\rm{kin}}}{m_{\rm{H}} \mu_{\rm{p}}}}
\end{equation}

\noindent
we have assumed T$_{\rm{kin}}$ = 15 K, consistently with \cite{cosentino2019} and mean molecular mass $\mu_{\rm{p}}$=2.37, for a typical interstellar elemental abundance. We find M$_{\rm{J}}$ in the range 0.3-0.7 M$_{\odot}$ and $\lambda_{\rm{J}}$$\sim$0.03 pc for all cores. We find that M/M$_{\rm{J}}>$2 and up to 20 for all 16 cores. We also investigate the cores spatial distribution, by calculating their minimum separation. For this we use the {\sc python} package {\sc networkx}\footnote{\url{https://pypi.org/project/networkx/}} to calculate the cores Minimum Spanning Tree (MST), that is the minimum set of edges connecting the core centre positions. We obtain that the minimum separation between cores is in the range 0.07-0.5 pc (average $\sim$0.2 pc), which is comparable to the Jeans length calculated for the ridge and the inner cloud and larger than the core Jeans length. This indicates that, if not further supported, for instance by turbulence and/or magnetic field, the cores are unstable against thermal support.\\

\noindent 
We now test the core stability, considering both thermal and turbulent support in the form of kinetic energy E$_{\rm{kin}}$ against gravity, in the form of potential energy E$_{\rm{pot}}$. The balance between this two terms leads to the so called virial parameter $\alpha_{\rm{vir}}$ \citep{bertoldi1992}. In the idealised case of a spherical core of uniform density supported by only kinetic energy (i.e. neglecting magnetic fields), the virial parameter takes the following form :

\begin{equation}
    \alpha_{\rm{vir}}=a\frac{5 \sigma_{\rm{tot}}^2 r}{GM}
\label{eq7}
\end{equation}

\noindent
where $\sigma_{\rm{tot}}$ is the N$_2$H$^+$ velocity dispersion including both thermal and non-thermal broadening (\citealp{fuller1992}; see also Eq. 6 in \citealp{barnes2023}). We note that Eq.~\ref{eq7} does not take into account contributions due to in-falling and/or out-flowing material. The factor $a$ accounts for systems with non-spherical and non-homogeneous density distributions and for most of these a value $a$=2$\pm$1 is assumed \citep{bertoldi1992}. Hence, cores with $\alpha_{\rm{vir}}<$2 are considered bound and unstable against collapse, while cores with $\alpha_{\rm{vir}}>$2 will likely be dispersed. We find that 20 out of 26 cores have $\alpha_{\rm{vir}}<$2, 4 cores have $\alpha_{\rm{vir}}\leq$4 and 2 cores have $\alpha_{\rm{vir}}>$4. Hence, 20 out of 26 cores are unstable to collapse when only supported by thermal and turbulent pressure. However, the uncertainty on $\alpha_{\rm{vir}}$ accounts for $\sim$40$\%$, with the major contribution due to the mass surface density uncertainty. Therefore, $\alpha_{\rm{vir}}\leq$4 are also consistent with the bound core scenario, within the uncertainties.\\

\noindent
We also test the core stability when support from magnetic field is included. In this case, the virial parameter takes the following form \citep{pillai2011}:

\begin{equation}
    \alpha_{\rm{B, vir}}=a \frac{5 r}{GM} \Big(\sigma_{\rm{tot}}^2 - \frac{v_{\rm{A}}^2}{6}\Big)
\end{equation}

\noindent
where the Alfvén velocity is $v_{\rm{A}}$= $B(\nu_{\rm{0}}\rho)$, with $B$ being the magnetic field strength and $\nu_{\rm{0}}$ is the permeability of free space. As for the previous case, no in-falling or out-falling material is taken into account. Hence, by setting $a$=2$\pm$1 and solving the equation for $B$, we infer that the required magnetic field pressure to support the cores against collapse is in the range 0.1$-$1 mG. For comparison, the magnetic field strength reported in literature are of the order of few mG in high-mass star forming regions \citep[e.g. ][]{beltran2019,dellolio2019}, in the range 0.2-5 mG \citep[e.g. the BISTRO survey; ][]{kwon2018} toward low-mass star forming regions and $\sim$0.5 mG in IRDCs \cite{pillai2015,soam2019}. These values are sufficient to support the cores against collapse.\\

\begin{table*}[t]
\caption{Properties of the identified cores.}
\centering
\label{tab:TabCores}
\begin{tabular}{lllllllllllll}
\hline
\hline
Idx & $x_{\rm{core}}$ & $y_{\rm{core}}$ &$\Sigma$ & Mass & $r$ &$\sigma_{\rm{tot}}$ & n$_{\rm{H_2}}$ & t$_{\rm{ff}}$ &M$_{\rm{J}}$ &  $\lambda_{\rm{J}}$ &$\alpha_{\rm{vir}}$ & B\\
    & (deg) & (deg)& (g cm$^{-2}$) & (M$_{\odot}$) & (pc) & (km s$^{-1}$) & (10$^5$ cm$^{-3}$) & (10$^4$ yr) & (M$_{\odot}$)   &(pc)  &  & (mG)\\
\hline
1 & 284.189 & 1.369 & 0.07 & 3.4 & 0.031 & 0.6 & 3.6 & 5.2 & 0.9 & 0.042 & 3.9 & 0.6 \\
4 & 284.176 & 1.374 & 0.01 & 4.6 & 0.029 & 0.7 & 6.0 & 4.0 & 0.7 & 0.032 & 3.4 & 0.8 \\
5 & 284.181 & 1.375 & 0.11 & 8.1 & 0.033 & 0.6 & 6.8 & 3.7 & 0.7 & 0.030 & 1.9 & 0.7 \\
10 & 284.183 & 1.379 & 0.14 & 5.1 & 0.025 & 0.3 & 9.7 & 3.1 & 0.6 & 0.025 & 0.5 & 0.5 \\
11 & 284.198 & 1.380 & 0.15 & 11.2 & 0.036 & 0.7 & 7.5 & 3.6 & 0.6 & 0.029 & 1.7 & 0.7 \\
12 & 284.188 & 1.379 & 0.14 & 4.0 & 0.021 & 0.2 & 14.1 & 2.6 & 0.5 & 0.021 & 0.3 & 0.7 \\
13 & 284.186 & 1.380 & 0.12 & 3.3 & 0.021 & 0.4 & 11.2 & 2.9 & 0.5 & 0.024 & 1.0 & 0.1 \\
16 & 284.188 & 1.385 & 0.16 & 20.2 & 0.052 & 0.4 & 4.3 & 4.7 & 0.9 & 0.038 & 0.6 & 0.4 \\
18 & 284.190 & 1.385 & 0.15 & 3.1 & 0.019 & 0.5 & 14.1 & 2.6 & 0.5 & 0.021 & 1.9 & 0.8 \\
19 & 284.179 & 1.387 & 0.07 & 1.6 & 0.019 & 0.6 & 7.4 & 3.6 & 0.7 & 0.029 & 4.8 & 0.8 \\
20 & 284.192 & 1.388 & 0.13 & 3.2 & 0.022 & 0.5 & 9.7 & 3.1 & 0.6 & 0.025 & 1.6 & 0.5 \\
22 & 284.190 & 1.389 & 0.16 & 6.7 & 0.027 & 0.5 & 10.5 & 3.0 & 0.5 & 0.024 & 1.2 & 0.4 \\
23 & 284.180 & 1.390 & 0.09 & 4.2 & 0.028 & 0.7 & 5.7 & 4.1 & 0.7 & 0.033 & 3.6 & 0.8 \\
25 & 284.183 & 1.391 & 0.08 & 2.5 & 0.022 & 0.5 & 7.5 & 3.6 & 0.6 & 0.029 & 2.3 & 0.6 \\
26 & 284.194 & 1.392 & 0.08 & 0.8 & 0.013 & 0.3 & 10.5 & 3.0 & 0.5 & 0.024 & 1.7 & 0.4 \\
29 & 284.181 & 1.393 & 0.09 & 2.0 & 0.019 & 0.4 & 8.5 & 3.3 & 0.6 & 0.027 & 1.9 & 0.5 \\
30 & 284.183 & 1.393 & 0.09 & 3.5 & 0.025 & 0.6 & 6.6 & 3.8 & 0.7 & 0.031 & 3.2 & 0.8 \\
32 & 284.190 & 1.394 & 0.10 & 1.1 & 0.014 & 0.6 & 11.5 & 2.9 & 0.5 & 0.023 & 6.0 & 1.1 \\
33 & 284.186 & 1.395 & 0.12 & 4.1 & 0.026 & 0.7 & 7.2 & 3.6 & 0.7 & 0.029 & 4.0 & 1.0 \\
35 & 284.185 & 1.399 & 0.14 & 6.4 & 0.026 & 0.6 & 11.7 & 2.9 & 0.5 & 0.023 & 1.8 & 0.8 \\
36 & 284.188 & 1.400 & 0.15 & 6.7 & 0.033 & 0.3 & 5.6 & 4.1 & 0.7 & 0.033 & 0.7 & 0.3 \\
37 & 284.190 & 1.402 & 0.17 & 6.0 & 0.025 & 0.6 & 11.7 & 2.8 & 0.5 & 0.023 & 1.6 & 0.7 \\
38 & 284.186 & 1.402 & 0.13 & 3.8 & 0.021 & 0.3 & 12.8 & 2.7 & 0.5 & 0.022 & 0.7 & 0.4 \\
39 & 284.185 & 1.402 & 0.11 & 0.6 & 0.009 & 0.3 & 24.2 & 2.0 & 0.4 & 0.016 & 1.5 & 0.5 \\
40 & 284.178 & 1.402 & 0.09 & 2.5 & 0.022 & 0.3 & 7.0 & 3.7 & 0.7 & 0.030 & 1.1 & 0.1 \\
41 & 284.184 & 1.403 & 0.08 & 1.1 & 0.016 & 0.3 & 7.8 & 3.5 & 0.6 & 0.028 & 1.9 & 0.4 \\
42 & 284.182 & 1.403 & 0.08 & 1.1 & 0.014 & 0.3 & 11.4  & 2.9 & 0.5 & 0.023 & 1.9 & 0.5 \\
\hline
\end{tabular}
\tablefoot{Columns are the index assigned by {\sc astrodendro}, the equatorial coordinates of the core centre, the core mass surface density, mass and radius radius, the total velocity dispersion from the N$_2$H$^+$ spectra, the Hydrogen volume density and the free-fall time, the Jeans mass and length, the virial parameter and the minimum magnetic field strength to prevent the core collapse.}
\end{table*}

\begin{figure}
    \centering
    \includegraphics[width=\linewidth]{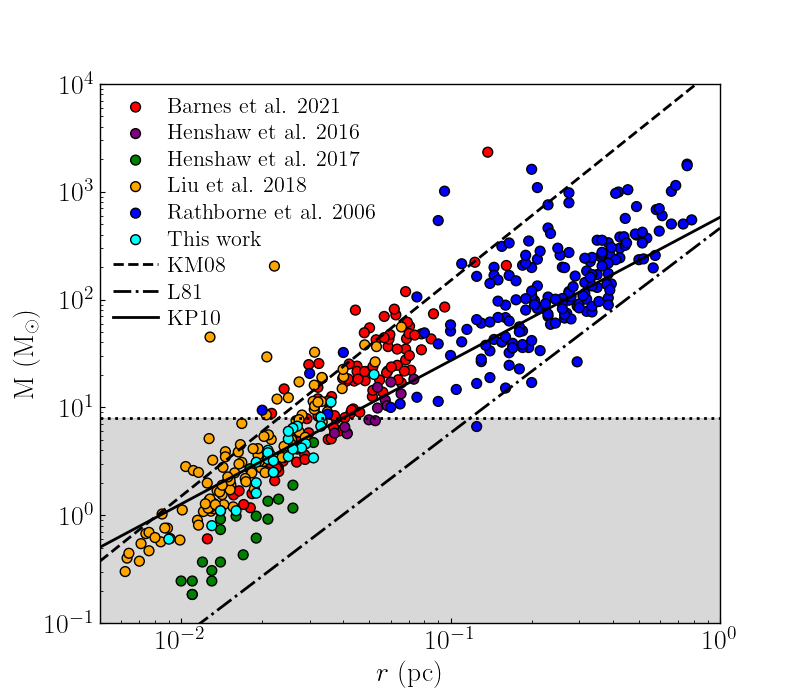}
    \caption{Core masses as a function of sizes for the cores identified in this this work (cyan dots). For comparison,  the mass and the sizes of the cores and/or clumps determined by \cite{rathborne2006,henshaw2016,henshaw2017,liu2018,barnes2021} and homogenised by \cite{barnes2021} are reported as coloured dots. Overlaid as the diagonal black lines are the mass–radius relations taken from \cite[][L81; dash-dotted line]{larson1981}, the high mass star formation thresholds taken from \cite[][KM08; dashed line]{krumholz2008} and \cite[][KP10; solid line]{kauffmann2010}. The KP10 relation has been scaled by a factor of 1.5 to match the dust opacity used in \cite{barnes2021}. The horizontal dotted black line shows the mass threshold of high-mass stars ($>$ 8 M$_{\odot}$).}
    \label{fig:MSrel}
\end{figure}

\noindent 
The IRDC G034.77 has been extensively studied as part of a larger samples. In \cite{rathborne2006}, the source was part of a sample of 38 IRDCs, while in more recent works \citep{henshaw2016,henshaw2017,liu2018,barnes2021} the cloud is part of a 10 IRDCs sample, first presented by \cite{butlerTan2009,butlerTan2012}. In Figure~\ref{fig:MSrel}, we compare the core mass-size relation here estimated with those reported by these authors in the larger samples and estimated from 1 and 3 mm continuum observations. For this, we used the homogenised sample presented by \cite{barnes2021}, that has been obtained by standardising the parameters used to determine the cores/clumps physical properties. 

\begin{figure}
    \centering
    \includegraphics[width=\linewidth]{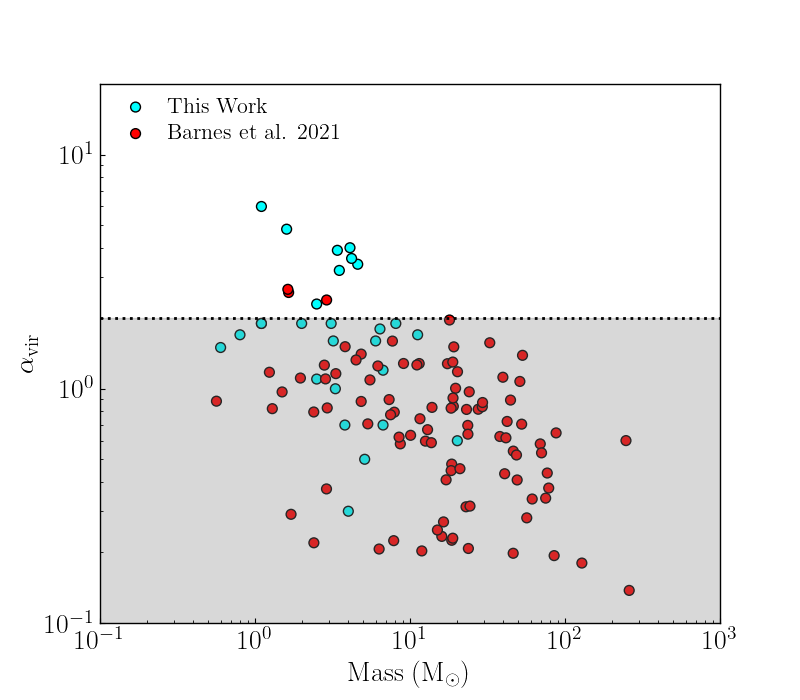}
    \caption{Core virial parameters as a function of core masses for the cores identified in this this work (cyan dots) and compared with the results from \cite{barnes2021} (red dots). The horizontal dotted black line and the grey area shows the virial parameter threshold for bound cores.}
    \label{fig:alphMass}
\end{figure}

\noindent
From Figure~\ref{fig:MSrel}, the cores identified in this work well follow the mass-size trend of the homogenised sample. The cores mostly lie on the low-mass end with just two exceptions, and sit on top of or below the KP10 relation. Hence, if the cores will harbour star formation, this may result in a population of low-mass stars. The few exception to this are the cores 11 and 16, which are the most extended ($r\sim$0.04-0.05 pc) and massive (11 and 20 M$_{\odot}$) of our sample and have virial parameter $<2$. Hence, these cores have the potential to host massive star formation assuming that no further fragmentation occurs and that the magnetic field is not strong enough to prevent their collapse. The mass-size relation of the G034.77 cores is in rough agreement (within the uncertainty) with the relation estimated by \cite{kauffmann2010}. In Figure~\ref{fig:alphMass}, we also compare the relation between virial parameter and core masses from this work with that inferred by \cite{barnes2021}. From Figure~\ref{fig:alphMass}, the two trends are well in agreement and largely overlap. In light of these comparisons, the properties of the cores here estimated follow well the trends estimated from a sample of sources that have similar masses, distances and contrast against the mid-IR Galactic background to G034.77 \citep{butlerTan2009, butlerTan2012}. 

\noindent

\section{Discussion and conclusions}\label{DiscAndConcs}
The mechanisms responsible for the ignition of star formation in molecular clouds are still under debate. Among the several scenarios, it has been proposed that large-scale interstellar shocks driven by stellar feedback, such as SNRs, may regulate star formation in galaxies \citep[e. g.][]{inutsuka2015,khullar2024}. By means of the interplay between positive and negative feedback, interstellar shocks may be able to set the low-levels of star formation efficiency typically measured in nearby galaxies. Several theoretical predictions have been proposed for this scenario, but poor observational constraints are available in literature. In particular, it is unclear whether SNRs can trigger star formation in impacted molecular clouds. In this work, we have studied the morphology and kinematics of the typical dense gas tracer N$_2$H$^+$ toward the IRDC G034.77, shock interacting with the SNR W44. The N$_2$H$^+$ emission is known to probe very dense and cold gas \citep[critical density 10$^{5}$ cm$^{-3}$;][]{jorgensen2004} and often used to identify cores in molecular clouds.\\
From the analysis presented in this work, the N$_2$H$^+$ emission is not widespread across the source, but it is mostly organised into two elongated structures i.e., one mainly distributed toward the ridge and the region named Clump, and one located further in (toward east) within the cloud. From the {\sc scousepy} analysis, the elongation toward the ridge has velocity in the range 38-41 km s$^{-1}$ while the eastern elongation splits into two spatially separated velocity components at 41-43 and 43-45.5 km s$^{-1}$, respectively. This higher-velocity component is consistent with the cloud systemic velocity, 43.5 km s$^{-1}$ as reported by \citep{simon2006,rathborne2006,hernandez2015}. Moreover, the more accurate {\sc scousepy} analysis shows that the low- and mid- velocity components spatially overlap toward the north-east of the cloud i.e., where the Clump is located, and kinetically connected not just toward the Clump but also marginally toward the centre of the innermost elongation. The velocities of these two components are consistent with the minimum velocities achieved in the J- and C-type shock respectively. Indeed, from the SiO map presented in \cite{cosentino2019}, the shock is approaching the ridge from a velocity of $\sim$46 km s$^{-1}$ and decelerates to 39 km s$^{-1}$ while plunging into the ridge. This kinematics is consistent with a CJ-type MHD shock, where the J-type component has velocities 46-42 km s$^{-1}$ and the C-type component has velocity 42-39 km s$^{-1}$. Hence, we speculate that the shock interaction driven by W44 may have penetrated within the cloud further than what showed by the SiO emission. Indeed, SiO emission is known to trace shocked gas that is very dense gas \citep[critical density $>$10$^5$ cm$^{-3}$;][]{schilke1997} and more in general it is a good probe of the most powerful parts of the shock. In our proposed scenario, the SiO emission only probed 'the tip of the iceberg' i.e., the most powerful part of a more extended shock. In line with this, since the strongest shock is occurring toward the ridge, the gas there is decelerated to lower velocities compared to the material within the inner cloud. Here, the much larger mass surface density (more than a factor of 2) may have prevented the gas to be decelerated as strongly as toward the ridge and helped the shock to be dissipated more quickly \citep{shima2017}. Finally, at this larger density the SiO freeze-out time is shorter than that at the ridge density \citep{martinpintado1992,jimenezserra2005}. A supporting factor to this scenario is the fact that both the ridge and the innermost elongation stand out as almost isolated features in the mass surface density map and have orientation similar to that of the SiO shock front. In particular, while the ridge appears as a spatially independent feature, the innermost elongation correspond to a large over-density within the cloud.\\ 

\noindent
While no data are available to estimate the D/H ratio for the innermost emission, we have confirmed that the N$_2$H$^+$ deuteration toward the ridge is significantly enhanced with respect to that measured toward the Ambient position. In \cite{cosentino2023}, we identified this position as probing the bulk of the cloud, for being far away from the shock front and devoid of clear signatures of ongoing star formation. However, from the high-resolution N$_2$H$^+$ image here investigated, the Ambient position is located in the vicinity of core 11 and not too far way form the high-velocity N$_2$H$^+$ component. Hence, in light of the new results, we cannot exclude that also the Ambient position may be affected by shocks and that these effects are not detectable within the relatively large beam of the single-dish observations. High-resolution N$_2$D$^+$ images are needed to further investigate this. With the data currently in hand, the Ambient position seems not associated neither with bright N$_2$H$^+$ emission nor with mass surface density peaks, as shown in Figures~\ref{fig:fig2} and ~\ref{fig:FigDendo}. This, together with the lack of N$_2$D$^+$(1-0) emission reported in \cite{cosentino2023}, hints toward the idea that the gas across the ridge is denser and colder than that toward the Ambient position. Indeed, high D/H ratios are known to be positively correlated with CO depletion \citep[e.g. ][]{miettinen2013,punanova2016,barnes2016}, which occurs for gas temperature below 20 K \cite{caselli1999}. We will present detailed investigating of the CO depletion in G034.77 in a series of forthcoming works (Petrova et al. in prep.; Segev et al. in prep.), using multiple CO transitions mapped with both the Green Bank 100 m antenna and ALMA. However, these preliminary studies seem to indicate relatively high CO depletion levels toward the ridge and the inner cloud ($\geq$4), but not the Ambient, suggesting indeed that the shock compressed gas is dense \citep[n($_{\rm{H_2}})\geq$10$^5$ cm$^{-3}$][]{cosentino2019} and cold.\\

\noindent
In \cite{cosentino2023}, the single pointing observations did not allow us to investigate the spatial morphology of this compressed material. Hence, we were unable to rule out whether this post-shocked gas is organised as a continuous slab of compressed material or if it has fragmented into cores. In the high-resolution ALMA images here analysed, the dense gas appear as highly fragmented, especially toward the ridge, with several core-like structures. These cores are well identified in our dendogram analysis and are spatially associated with well emission peaks in the mass surface density map. The cores have masses in the range 1-20 M$_{\odot}$ and well follow the mass-size trend of the homogenised core sample presented by \cite{barnes2021} that also includes results from \cite{rathborne2006,henshaw2016,henshaw2017,liu2018} as well as literature works \citep{larson1981,sanchezMonge2013,peretto2013,kauffmann2013,kainulainen2013,chen2020}. We suggest that these cores do not host deeply embedded protostars but may either be in a starless or pre-stellar stage for the following reasons.  Most cores appear as density enhancements in  the mass surface density map from \cite{kainulainen2013} (see Figure~\ref{fig:FigDendo}). Moreover, the Herschel 70 $\mu$m image in Figure~\ref{fig:Mosaic} shows no significant emission peaks associated with the cores. No molecular outflow signatures are found in our current SiO ALMA map \citep{cosentino2019}, and this seems to be confirmed by the CO ALMA emission maps that will be presented in Segev et al. (in prep). All this suggests that the cores do not host deeply embedded protostars and thus the identified cores may either be in a starless or pre-stellar stage. None of the cores here identified is detected in the 1mm and 3 mm continuum images of \cite{rathborne2006} and \cite{barnes2021,cosentino2019}, respectively. However, it is important to note that N$_2$H$^+$ is a resilient tracer of imminent star formation in molecular clouds \citep{priestley2023} and that starless cores with no mm continuum emission have been reported by means of N$_2$D$^+$ emission \citep[e.g., ][]{kong2017}. Additionally, since the continuum emission is the most established method to identify cores in molecular clouds, our current understanding of the correlation between starless core and mm continuum emission could be biased. Finally, we note that the detection limit of the \cite{rathborne2006}  single-dish 1.2 mm maps, due to beam size and source distances effects, is $<$45 m$_{\odot}$ well above the core masses here reported. Higher-resolution 1 mm maps are therefore needed to establish whether the cores are associated with continuum sources.\\ The minimum separation between cores is, on average, 0.2 pc and consistent with the fragmentation scales of the inner cloud and the ridge. Moreover, except for a few, the cores appear unstable against gravitational collapse in the presence of thermal and turbulent pressure. We estimate that a magnetic field 0.5-1 mG may support these cores against collapse. These values are consistent with the $\sim$0.85 mG magnetic field indirectly estimated toward G034.77 in \cite{cosentino2019}, but this value is relatively uncertain since it was indirectly inferred from shock models.\\  

\noindent
The field of view of our N$_2$H$^+$ ALMA map does not cover the region of the cloud nearby the H{\small II} region, where \cite{paron2009} reported evidence of star forming cores. These sites are associated with molecular outflows and point-like IR emission and are likely more evolved than the cores here identified. Furthermore, as reported by the authors, they are spatially aligned with the expanding H{\small II} region shells. As consequence, the authors conclude that star formation was not triggered by the SNR but rather by the H{\small II} region through the collect and collapse process. On the other hand, the field of view of our observations is well covered by the IR and molecular tracers maps of \cite{paron2009}, but the authors do not report evidence of relatively evolved star formation activity in the vicinity of W44. Hence, it is unlikely that the cores here identified have been produced by the H{\small II} region and/or pre-SNR stellar wind compression. This indicates that, if the cores here identified will form stars, the two shock compression i.e., the H{\small II} region and the SNR-driven shocks, will lead to star populations at different evolutionary stages i.e., the multi-episodic star formation often observed in molecular clouds \citep[e.g.,][]{beccari2019}. \\

\noindent
In light of all this, we suggest a scenario in which the large-scale shock driven by the SNR W44 has penetrated within the IRDC G034.77, compressing the gas along both a high-density elongation within the cloud and a lower density ridge at the edge. Within these features, the shock passage may have triggered the formation of cores at spatial scales consistent with the Jeans length. Due to the lack of deeply embedded sources and molecular outflows signatures, the cores are either in a starless or pre-stellar stage and appear stable against gravitational collapse, although the importance of magnetic field needs to be further addressed. Further high-resolution observations of dedicated tracers are needed to confirm this scenario.   

\begin{acknowledgements}
This paper makes use of the following ALMA data: ADS/JAO.ALMA\#2017.1.00687 and ADS/JAO.ALMA\#2018.1 .00850. ALMA is a partnership of ESO (representing its member states), NSF (USA) and NINS (Japan), together with NRC (Canada), NSTC and ASIAA (Taiwan), and KASI (Republic of Korea), in cooperation with the Republic of Chile. The Joint ALMA Observatory is operated by ESO, AUI/NRAO and NAOJ. This work is based on observations carried out under project number 028-20 with the IRAM 30m telescope. IRAM is supported by INSU/CNRS (France), MPG (Germany) and IGN (Spain). This publication is based on data acquired with the Atacama Pathfinder Experiment (APEX) under programme ID 0111.F-9303. APEX is a collaboration between the Max-Planck-Institut fur Radioastronomie, the European Southern Observatory, and the Onsala Space Observatory. Swedish observations on APEX are supported through Swedish Research Council grant No 2017-00648. G.C., A.T.B. and M.D.S. acknowledge support from the ESO Fellowship Program. G.C. also acknowledges funding from the Swedish Research Council (VR Grant; Project: 2021-05589). J.C.T. acknowledges support from ERC project 788829–MSTAR. I.J-.S acknowledges funding from grant No. PID2019-105552RB-C41 awarded by the Spanish Ministry of Science and Innovation/State Agency of Research MCIN/AEI/10.13039/501100011033. J.D.H. gratefully acknowledges financial support from the Royal Society (University Research Fellowship; URF/R1/221620). R.F. acknowledges support from the grants Juan de la Cierva FJC2021-046802-I, PID2020-114461GB-I00 and CEX2021-001131-S funded by MCIN/AEI/ 10.13039/501100011033 and by “European Union NextGenerationEU/PRTR”. S.V. acknowledges partial funding from the European Research Council (ERC) Advanced Grant MOPPEX 833460. S.V. and J.C.T acknowledge the support from a Royal Society International Exchanges Scheme grant (IES/R3/170325). 
\end{acknowledgements}

%
\bibliographystyle{aa} 
\bibliography{aa.bib} 
%
\end{document}